\documentclass[12pt]{article}
\usepackage{amsmath}
\usepackage{amssymb}
\usepackage{graphicx}
\usepackage{enumerate}
\usepackage{natbib}
\usepackage{tabularx}
\usepackage{float}
\usepackage{url} 
\usepackage{accents}
\newcommand{\blind}{1}
\usepackage[font=small,skip=0pt]{caption}
\usepackage[ruled]{algorithm2e}
\usepackage{hyperref}

\usepackage{array}
\usepackage{multirow}
\usepackage[noblocks]{authblk}

\begin{document}

\def\spacingset#1{\renewcommand{\baselinestretch}%
{#1}\small\normalsize} \spacingset{1}

\if1\blind

\title{Studying Competing Events with Federated Cumulative Incidence Curves}

\author[1,c]{Malcolm Risk}
\author[2]{Shuang Yang}
\author[3]{Jiang Bian} % If author has multiple affiliations
\author[2]{Yi Guo}
\author[4]{Hyojung Jang}
\author[5]{Serena Guo}
\author[1]{Xu Shi}
\author[4]{Lili Zhao}

\affil[1]{Dept of Biostatistics, University of Michigan}
\affil[2]{Dept of Health Outcomes and Biomedical Informatics, University of Florida}
\affil[3]{Dept of Biostatistics \& Health Data Science, University of Indiana}
\affil[4]{Dept of Preventive Medicine, Northwestern University}
\affil[5]{Dept of Pharmacy Practice, Purdue University College of Pharmacy}
\affil[c]{Corresponding Author: mhrisk@umich.edu}
  \date{}

 \maketitle

\if0\blind
{
  \begin{center}
    {\large \bf Federated Statistical Inference for Competing Risk Curves and Restricted Mean Time Lost}
\end{center}
} \fi
\vspace{-0.8cm}
\begin{abstract}
\footnotesize \noindent \textbf{Objective}: Combining electronic health record (EHR) data from multiple institutions is a valuable strategy for conducting post-market safety surveillance of medical products, but privacy concerns limit sharing individual-level data. We develop a novel federated learning (FL) method for multi-site post-market safety surveillance of medical products using competing risks data. We apply this novel method to study immune-related adverse events (irAEs) following treatment with immune checkpoint inhibitors (ICIs) in patients with pre-existing auto-immune disease (AID).  \\ \textbf{Methods}:   We provide an algorithm for constructing non-parametric cumulative incidence curves for competing event types, which can be used to compare exposure groups (e.g. treated and untreated) with no sharing of patient-level data across institutions. We incorporate covariate adjustment via inverse propensity weighting, and informative causal comparison using the area under cumulative incidence curves, known as restricted mean time lost. \\ \textbf{Results}: We apply our method to $N=10,281$ cancer patients with no pre-existing endocrine-related AID receiving ICIs across $K=10$ sites from the OneFlorida+ network, comparing patients with a pre-existing non-endocrine AID to those with no pre-existing AID. After covariate adjustment, we found that patients with a pre-existing non-endocrine AID lost 4.8 [95\% CI: 4.3,5.2] months of event-free survival time to endocrine irAEs in the first 18 months following treatment, compared to 3.2 [95\% CI: 3.1,3.3] months in the group without prior AID. \\ \textbf{Conclusion}: As patients with prior AID were initially excluded from clinical trials of ICIs, our findings provide important new information to clinicians and patients receiving or considering ICI treatment. Our proposed non-parametric federated algorithm is the first to allow investigators to use some of the most crucial non-parametric tools for conducting postmarket safety surveillance across multiple institutions.
\end{abstract}

\noindent%
\vfill

\newpage
\spacingset{1.9} 
\section{Introduction}
\label{sec:intro}

Routinely collected patient data from electronic health records (EHR) have become a crucial source of data for conducting post-market safety surveillance of medical products. EHR systems can provide more accurate and rapid detection of safety problems compared to phase IV clinical trials or public reporting systems \citep{Nelson2015}. To achieve larger sample size and enhanced generalisability, researchers have gravitated towards large federated hospital networks to analyse EHR data from multiple sources \citep{Fleurence2014, Hripcsak2015, Verma2021}. Since privacy concerns and strict approval processes limit the sharing of individual-level data, federated learning (FL) methods have been developed to fit statistical and machine learning models across many institutions without sharing patient data. The federated approach is particularly useful when pre-licensure clinical trials provide limited information, such as when we expect a higher risk of adverse events in certain patient subgroups. One such case is the use of immune checkpoint inhibitors (ICIs) to treat cancer in patients with pre-existing auto-immune disease (AID). As ICIs can trigger immune-related adverse events (irAEs), patients with AID were initially excluded from ICI trials. Over time, a consensus has emerged that ICIs should still be offered to patients with AID, necessitating better evidence regarding the risk of irAEs in this population \citep{Chen2019, Khandwala2025, Capelli2022}. As with many other types of adverse events, irAEs are time-to-event outcomes whose occurrence is precluded by other competing events such as death. Censoring patient follow-up at competing events can introduce bias or confusing interpretation \citep{Young2020}, so timely and accurate evidence requires federated methods for studying competing risks.

Outside the federated context, there are a wide variety of popular approaches for studying competing risks data. There is a family of non-parametric methods, including the estimation of the cumulative incidence function (CIF) for each event type \citep{Peterson1975}; test statistics for comparing curves for different groups of patients \citep{Lin1997}; and the estimation of area under the curve, known as restricted mean time lost (RMTL) \citep{Conner2021}. There is the semi-parametric Fine-Gray model, which generalizes Cox regression models to the competing risk setting \citep{Fine1999}, and a less popular set of parametric regression models \citep{Hinchliffe2013}. Finally, there are modern approaches where each observation is converted into a continuous ``pseudo-observation" based on the change in the
estimated cumulative incidence curve caused by removing the observation, with the pseudo-observations then used as outcomes in flexible regression or machine learning models \citep{Binder2014}. 

Existing work on federated competing risks analysis is exclusively focused on the semi-parametric and pseudo-observation approaches. For semi-parametric approaches, \cite{Zhang2024} have proposed ODACoR, which is a one-shot federated algorithm implementing the Fine-Gray model. This approach allows investigators to obtain hazard ratio estimates for the effect of covariates on events of each specific type. \cite{Rahman2023CR} have proposed using pseudo-observations to train federated machine learning models. Their method requires obtaining a global risk table through obtaining a local risk table from each site (number at-risk and with an event at unique event times), which is used to calculate pseudo-observations to use as outcomes in a machine learning model. This allows for conducting flexible risk prediction, but does introduce some privacy risk as individual patient survival times and event indicators are easily identifiable from local risk tables. 

The goal of this paper is to extend non-parametric competing risks methods to the federated context. We believe that this provides a valuable alternative to existing methods, for a number of reasons. Compared to the pseudo-observation methods, we eliminate the potential risk of leaking individual patient data by avoiding sharing local risk tables. In addition, we provide intuitive and causally meaningful results that are easily interpreted by clinicians and regulators, which are often preferred to black-box predictions for high-level policy decision-making. Compared to the semi-parametric Fine-Gray model, we avoid the proportional hazards assumption, which is especially difficult to test or relax in the multi-center context. Non-parametric methods also provide highly intuitive visualizations in the form of cumulative incidence curves, which can prove a crucial complement to regression-based methods when communicating results to a broader audience. 

In this paper, we present a novel federated method for non-parametric estimation of the CIF and RMTL, with accompanying covariate adjustment using inverse propensity weighting (IPW) \citep{Austin2025}. We use simulated data to demonstrate that our method can achieve the same performance as the equivalent analysis with no data sharing constraint. We then apply our method to data from the OneFlorida+ research network \citep{Hogan2022} to conduct an impactful real-world study comparing irAE incidence following ICI treatment in patients with pre-existing AID to patients without pre-existing AID. The R code used to implement our approach, along with detailed documentation, can be downloaded at https://github.com/MR236/FederatedCR.

\section{Methods}
\label{sec:meth}

\subsection{Data Structure and Notation}

The motivating setting for this paper is studying the impact of a binary exposure, pre-existing auto-immune disease (AID), on two competing event types: death and immune-related adverse event (irAE). In an ideal study where patients were not subject to censoring, we would observe the data $(T_i,\Omega_i, A_i, \mathbf{Z}_i)$ for each subject $i$. Exposure status is given by $A_i=1$ for pre-existing AID and $A_i = 0$ for no prior AID, event type is given by $\Omega_i=1$ for irAE and $\Omega_i=2$ for death, $T_i$ represents time from immune-checkpoint inhibitor (ICI) treatment until event, and $\mathbf{Z}_i$ are confounding variables. In reality, due to loss to follow-up, we will only observe data for each individual up to a censoring time $C_i$. We therefore denote the observed data as $(X_i,\Delta_i, A_i, \mathbf{Z}_i)$, where the follow-up time $X_i=\min(T_i, C_i)$ represents the amount of time that subject $i$ remains in the study and the event status $\Delta_i = I(T_i < C_i)\Omega_i$ takes value $\Delta_i = 0$ if a patient is censored, $\Delta_i = 1$ for an irAE, and $\Delta_i = 2$ for death. Although our method allows for an arbitrary number of competing event types, we limit our description to the case of two potential types or ``causes" for simplicity.

\subsection{The Cumulative Incidence Function}

In a typical survival analysis with only one event type, we would be interested in the counterfactual survival function $S^a(t)=P(T_i^{A_i=a} > t)$, which represents the probability of remaining event-free up to $t$, given that all individuals receive exposure level $a$. We might also be interested in the cumulative incidence function (CIF), $F^a(t)=1-S^a(t)$, representing the complementary probability of having an event prior to $t$.

In the context of competing risks data, we have a version of the cumulative incidence function that focuses on events of a particular type, or ``cause". To study the impact of having a pre-existing AID ($A_i = 1$) on incidence of endocrine irAEs (event type $\Omega_i=1$), we want to estimate the counterfactual cause-specific CIF:
\begin{equation}\label{eq:CIF}
F_k^a(t) = P(T_i^{A_i=a} < t, \Omega_i^{A_i=a} = k)
\end{equation}
where $T_i^{A_i=a}$ and $\Omega_i^{A_i=a}$ represent the values $T_i$ and $\Omega_i$ would have taken if individual $i$ was in exposure level $a$. Our goal is to estimate the counterfactual CIF for the exposed group $F_1^1(t)$ and the equivalent curve in the control group $F_1^0(t)$. The comparison between these curves will indicate the impact of AID on irAE risk after removing the effect of other covariates.  

The sum of the cause specific CIFs across all event types is equal to the overall CIF, giving a relationship between the survival function and the CIFs: $S^a(t)=1-\sum_{k=1}^K F_k^a(t)$. This means that first estimating the survival function can be used to make the process of estimating the cause-specific CIFs easier. In previous work \citep{Risk2025}, we developed a federated weighted Kaplan-Meier (KM) estimation method to obtain estimates $\hat{S}^1(t)$ and $\hat{S}^0(t)$ for the counterfactual survival curves in the exposed and control arms, respectively; this method is a key part of the federated estimation process described in this paper.

\subsection{Inverse Propensity Weighting}

 In observational settings characteristics of exposed patients ($A_i=1$) will differ from those of control patients ($A_i=0$), and so the observed outcomes in each group will not be representative of the overall population. To account for these covariate imbalances, we use inverse propensity weighting (IPW). We start by fitting a propensity score model, defined by $g(P(A_i=1\mid \mathbf{Z_i}))=\boldsymbol{\alpha}^T\mathbf{Z_i}$, where $g(\cdot)$ is typically the logit link function. Then we assign weights to each individual based on their predicted probability of receiving their actual exposure assignment, $\hat{w}_i = \frac{1}{\hat{p}_i}$ for treated subjects and $\hat{w}_i = \frac{1}{1-\hat{p}_i}$ for untreated subjects. After the application of weights, we will have treated and untreated pseudo-populations with balanced characteristics. 
 
\subsection{CIF Estimation in a Single-Centre Study}

Estimating the counterfactual cumulative incidence for a particular cause requires an estimate for the counterfactual survival function for each group, $S^a(t)=P(T_i^{A_i=a} < t)$, which represents the probability of an individual experiencing no event (of any type) prior to time $t$, assuming that they received exposure $a$. In a single-center study, we can obtain such an estimate through the weighted Kaplan-Meier (KM) method \citep{Xie2005}. Once the weights and corresponding weighted KM curves have been obtained, we can estimate the counterfactual CIFs \eqref{eq:CIF} using the following non-parametric estimator \citep{Kalbfleisch1980}:
\begin{equation}\label{eq:CR}
\hat{F}_k^a(t) = \sum_{j|t_j \leq t} \hat{S}^a(t_{j-1}) \frac{d_{kj}(a)}{n_j(a)}
\end{equation}
where $\{t_j:t_j \leq t\}$ is the set of unique event times observed prior to time of interest $t$, $d_{kj}(a) = \sum w_i I(A_i=a)I(\Delta_i=k) I(X_i=t_j)$ is the weighted number of events of type $k$ at $t_j$ in exposure group $a$, and $n_{j}(a) = \sum w_i I(A_i=a)I(X_i>t_j)$ is the weighted number of individuals at-risk at $t_j$ in exposure group $a$. 

Under the assumptions of exchangeability, positivity, and consistency described in \cite{Young2020}, this weighted estimator will identify the counterfactual CIF \eqref{eq:CIF}. In a multi-center study, equation \eqref{eq:CR} can be calculated only if we are able to share individual-level data, and hence the alternative approach described in the forthcoming sections.

\subsection{CIF Estimation in a Multi-Centre Study}

\begin{figure*}[!t]
\centering
{\includegraphics[width=1.02\textwidth]{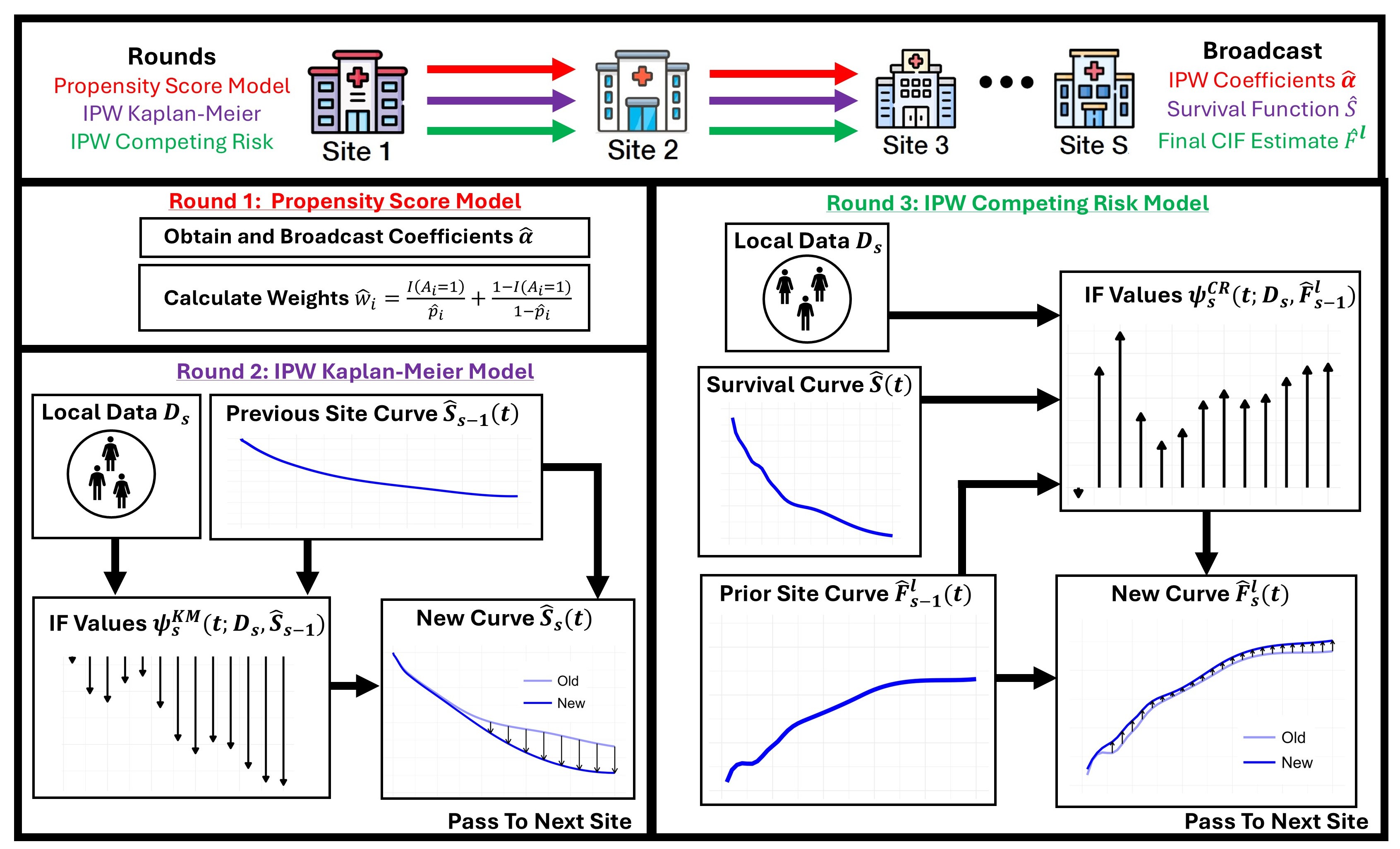}}
\caption{Schematic of update steps for proposed federated estimation of the CIF. \label{fig:first}}
\end{figure*}

Our approach for multi-center estimation of the CIF involves three rounds of sequential communication between the participating sites. Figure \ref{fig:first} shows the construction of our federated CIF estimation algorithm, while Algorithm~\ref{algorithm:code_DKM} gives mathematical details. Our method is sequential rather than centralized, with each site sharing updated summary statistics with only the next site in a prespecified order. Within each round, we recommend ordering the sites from the largest to the smallest sample size, which maximizes accuracy by increasing the rate of convergence of the estimates. The first two rounds rely on existing approaches to estimate the weights and counterfactual survival functions, which are then used in the third round to estimate the CIF. In the first round, we calculate the coefficients $\boldsymbol{\alpha}$ of the propensity score model using any federated estimation method for GLMs, such as renewable estimation \citep{Luo2020} or ODAL \citep{Duan2020}. These coefficients are broadcast to all sites, who can use them to calculate a weight $\hat{w}_i=\frac{A_i}{\hat{p_i}}+\frac{1-A_i}{1-\hat{p_i}}$, where $\hat{p_i} = g^{-1}(\boldsymbol{\alpha}^T\mathbf{Z_i})$. In the second round, we use our previously published algorithm for federated KM estimation \citep{Risk2025} to obtain estimates $\hat{S}^1(t)$ and $\hat{S}^0(t)$ for the counterfactual survival curves in the exposed and control arms, respectively. This algorithm also allows us to obtain estimates $\hat{Y}^1(t)$ and $\hat{Y}^0(t)$ for the at-risk functions, where $Y^a(t)=P(X_i^{A_i=a} > t)$ is the probability an individual remains at-risk at time point $t$ given exposure level $a$. In order to protect patient privacy, these estimates are passed between sites in the form of a continuous, monotonic polynomial spline function rather than the typical step function where we might run the risk of leaking exact patient survival times. Our previous work contains extensive simulations showing that this method can achieve identical efficiency and accuracy to the weighted KM method in the pooled data.

Once we have obtained the estimated weights, survival functions, and at-risk functions, we obtain CIF estimates for each event type and exposure level using a novel federated approach in a single round of sequential communication. The core of our federated approach is to begin with an initial estimate for the CIF, and have each site use their local data to apply an incremental update to the CIF such that we achieve accurate estimation without sharing individual-level data. To achieve this, we make use of the empirical \textit{influence function}, a statistical quantity that reflects the impact of adding an observation to the CIF estimate. In Appendix A we provide a detailed derivation and expression for the empirical influence function of the CIF. In our federated algorithm, we use a plug-in estimate for the empirical influence function of the CIF for an observation $o_i=(X_i, \Delta_i, \hat{w}_i)$ and estimates $\hat{S}^a, \hat{Y}^a, \hat{F}^a_k$: 
\begin{equation}\label{eq:INF}
\begin{split}
\psi(\hat{F}^a_k(t);o_i) &= \frac{\hat{w}_iI(\Delta_i=k) I(X_i \leq t) \hat{S}^a(X_i)}{\hat{Y}^a(X_i)} + \sum_{t_j \leq t}\frac{\hat{F}^a_k(t_j)- \hat{F}^a_k(t_{j-1})}{\hat{S}^a(t_{j-1})}\phi(\hat{S}^a(t_j);o_i)\\&- \hat{w}_i\sum_{t_j \leq (t \wedge X_i)} \frac{\hat{F}^a_k(t_j)- \hat{F}^a_k(t_{j-1})}{\hat{Y}^a(t_j)}
\end{split}
\end{equation}
where above $\phi(\hat{S}^a(t_j);o_i)$ is the empirical influence function of the counterfactual survival curve estimate $\hat{S}^a$ \citep{Risk2025} and $\{t_j:t_j \leq t\}$ is the set of unique event times observed prior to time of interest $t$. The empirical influence function is useful for federated estimation because it has similar properties to the score function of a parametric estimator; the estimate $\hat{F}^a_k$ from equation \eqref{eq:CR} solves the estimating equation $\sum \psi(\hat{F}^a_k(t);o_i) = 0$ for all unique event times $t \in \{t_1,\dots,t_J\}$. When deriving the empirical influence function, we assumed that weights were known, even though in practice they are estimated. This can lead to slightly over-conservative statistical inference, which we discuss in greater detail when describing our inference procedures.

To understand how our method can be used to estimate the CIF in a federated context, consider a scenario in which we have data from exposure level $a$ ($A_i = a$) divided into two sites, denoting data from site 1 as $D_1$, and data from site 2 as $D_2$, with sample sizes $N_1$ and $N_2$, respectively. We know that the estimate based on pooling the full data $\hat{F}^a_k(t; D_1 \cup D_2)$ will solve the following equation for all times $t \in \{t_1, \dots, t_J\}$: 
\begin{equation}\label{eq:EE}
\begin{split}
\sum_{D_1} \psi(\tilde{F}^a_k(t);o_i) + \sum_{D_2} \psi(\tilde{F}^a_k(t);o_i) = 0
\end{split}
\end{equation}
However, each site can only access its own local data. To solve the equation despite this constraint, we can approximate the first sum (see Appendix A for details) by using the site 1 estimate $\hat{F}^{a}_k(t; D_1)$:
\begin{equation}\label{eq:EE2}
\begin{split}
\sum_{D_1} \psi(\tilde{F}^a_k(t);o_i) \approx N_1 [\hat{F}^{a}_k(t; D_1) - \tilde{F}^a_k(t)]
\end{split}
\end{equation}
This means that site 1 can obtain an initial estimate $\hat{F}^{a}_k(t; D_1)$ for each $t \in \{t_1,\dots,t_J\}$ using \eqref{eq:CIF}, and pass this estimate to site 2. Then site 2 needs to solve the following equation, involving only local data $D_2$ and the prior estimate $\hat{F}^{a}_k(t;D_1)$:
\begin{equation}\label{eq:EE3}
\begin{split}
N_1 [\hat{F}^{a}_k(t;D_1) - \tilde{F}^a_k(t)] + \sum_{D_2} \psi(\tilde{F}^a_k(t);o_i) = 0
\end{split}
\end{equation}
For each value $t \in \{t_1,\dots,t_J\}$, this equation can be forward-solved using a version of Newton's method with a modified denominator to minimize the required communication between sites: 
\begin{equation}\label{eq:EE4}
\begin{split}
\hat{F}_k^a(t)^{(r+1)} = \hat{F}^a_k(t)^{(r)} + \frac{N_1 [\hat{F}^{a}_k(t;D_1) - \hat{F}^a_k(t)^{(r)}] + \sum_{D_2} \psi({\hat{F}_{k}^a}(t)^{(r)};o_i)}{N_1+N_2}
\end{split}
\end{equation}
In Appendix A, we show that $\frac{1}{N} \sum_{i=1}^N  \frac{\partial \psi(\hat{F}^a_k(t);o_i)}{\partial \hat{F}^a_k(t)} \overset{p}{\to} -1$, meaning that the denominator of our update step is asymptotically equivalent with large sample size to the denominator seen in Newton's method. This will produce an estimate $\hat{F}^{a}_k(t; D_1 \cup D_2)$ incorporating data from both sites. In our federated approach, we apply the same algorithm to update an estimate $\hat{F}^{a}_k(t; \tilde{D}_s)$ based on cumulative data from the first $s$ sites $\tilde{D}_s = \cup_{u=1}^{s} D_s$ to a new estimate $\hat{F}^{a}_k(t; \tilde{D}_{s+1})$ using data from site $s+1$.  

For the first step in the estimation round, the initial site uses estimator \eqref{eq:CR} to obtain an initial estimate based on it's local data. Then, to avoid risk to patient privacy, the site converts the curve to a continuous non-decreasing spline function $\hat{F}_k^a(t; \Theta_1)$ where $\Theta_1$ contains spline coefficients, degrees of freedom, and knot locations. These parameters are passed to the second site. Each subsequent site in the prespecified order follows an identical procedure to update the parameters $\Theta_{s-1} \to \Theta_{s}$, which are then passed forward until all sites have been incorporated. At each site, we shift the previous CIF estimates based on local data $D_s$ using the following iterative updating scheme, which is a generalization of equation \eqref{eq:EE4}, at a dense set of $Q$ time points $t \in \{t_1,\dots,t_Q\}$:
\begin{equation}\label{eq:EEMain}
\begin{split}
\hat{F}_k^a(t)^{(r+1)} = \hat{F}^a_k(t)^{(r)} + \frac{\tilde{N}_{s-1} [\hat{F}^a_k(t; \tilde{D}_{s-1})- \hat{F}^a_k(t)^{(r)}] + \sum_{D_{s}} \psi({\hat{F}_{k}^a(t)}^{(r)};o_i)}{\tilde{N}_{s-1} +N_s}
\end{split}
\end{equation}
A larger number of time points $Q$ leads to a more flexible and accurate set of updates to the existing curve. As the computational cost of this step is low, we are able to pick a very large number of time points (e.g. $Q=100$) evenly distributed across the follow-up time to maximize accuracy, although $15-20$ time points is typically sufficient. Algorithm convergence is assessed based on when the estimated cumulative sum of influence function values drops below a small value $\epsilon$.  
\begin{equation}\label{eq:EEConv}
\begin{split}
\left|\tilde{N}_{s-1} [\hat{F}^a_k(t; \tilde{D}_{s-1})- \hat{F}^a_k(t)^{(r)}] + \sum_{D_{s}} \psi({\hat{F}_{k}^a(t)}^{(r)};o_i)\right| < \epsilon
\end{split}
\end{equation}
Even for very small choices of epsilon (e.g. $\epsilon = 10^{-6}$) the updating scheme typically converges in less than 10 iterations and only a few seconds of computation time, even for sites with a large sample size. 

Based on the obtained $1\times Q$ vector of new CIF estimates $\hat{F}_k^a(\mathbf{t};\Theta_{s-1}, \mathbf{D}_{s})$, the site obtains a new set of spline parameters $\Theta_{s}$ and passes them to the next site.  At the final site, we will obtain our counterfactual CIF estimates: $\hat{F}_k^{1}(\mathbf{t})$ under treatment, and $\hat{F}_k^{0}(\mathbf{t})$ under no treatment. Comparing these two CIFs allows investigators to evaluate the effect of the exposure on the event type of interest. Our method require the first site in the order to be of reasonable sample size with $>10$ patients at-risk throughout the follow-up in each group being a reasonable rule of thumb. A major advantage of our method is that it can accommodate sites with as little as one observation, whereas many other methods (such as meta-analysis) require sufficient sample size to obtain a stable site-specific estimate for a site to contribute to the overall estimate.

\begin{algorithm*}[h]
\caption{\label{algorithm:code_DKM} \small Main algorithm for updating the CIF site-by-site.}
 \small
	\textbf{Inputs:} Local competing risk data $D_1, \dots, D_S$ across $S$ sites with sample sizes $n_1,...,n_S$. \\
	\textbf{Outputs:} Final CIF estimate $\hat{F}^a_k$ for exposure level $a$ and cause $k$. \par\smallskip
1. Fit a federated propensity score model using an existing method \citep{Luo2020, Duan2020} to obtain coefficients $\boldsymbol{\alpha}$ and treated fraction $\hat{p}$ and broadcast to all sites, who calculate a normalized weight $\hat{w}_i$ for each subject.  
\par
2. Obtain spline-based estimates for the survival $\hat{S}^a$ and at-risk functions $\hat{Y}^a$ using existing federated methods \citep{Risk2025} and broadcast to all sites.
\par
3. At site 1, obtain a spline-based estimate of the cumulative incidence function $\hat{F}^a_k(\cdot ;D_1)$ for exposure level $a$ and cause $k$. Select $P$ time points $t_1,\dots,t_P$ to conduct statistical inference and calculate squared influence function sums $v(t_p; D_1) = \sum_{D_1} \psi({\hat{F}_{k}^a(t_p)};o_i)^2$.   Pass spline parameters and influence function sums to site 2.
\par
4. Denote the cumulative data $\tilde{D}_s = \cup_{u=1}^{s} D_s$ with cumulative sample size $\tilde{N}_s$. For each site $s=2,\dots,S$ conduct the following steps: \par 
\quad \quad \textbf{Initialization}. Set initial value equal to the prior estimate, $\hat{F}^a_k(t)^{(0)}=\hat{F}^a_k(t; \tilde{D}_{s-1})$. \par
\quad \quad \textbf{Estimation}. Use the following iterative updating scheme to obtain $\hat{F}^a_k(t; \tilde{D}_{s})$:
\begin{equation*}\label{eq:EEAlg}
\begin{split}
\hat{F}_k^a(t)^{(r+1)} = \hat{F}^a_k(t)^{(r)} + \frac{\tilde{N}_{s-1} [\hat{F}^a_k(t; \tilde{D}_{s-1})- \hat{F}^a_k(t)^{(r)}] + \sum_{D_{s}} \psi({\hat{F}_{k}^a(t)}^{(r)};o_i)}{\tilde{N}_{s-1} +N_s}
\end{split}
\end{equation*}
 \par
 \quad \quad \textbf{Inference}. For each $t_p$, update squared influence function sums: 
\begin{equation*}\label{eq:EEAlgIF}
\begin{split}
v(t_p; \tilde{D}_{s}) = v(t_p; \tilde{D}_{s-1}) + \sum_{D_s} \psi({\hat{F}_{k}^a(t_p)};o_i)^2 \end{split}
\end{equation*}\par 
\quad \quad \textbf{Communication}. Pass $\hat{F}^a_k(t; \tilde{D}_{s})$ to site $s+1$. \par
5. At the final site $S$, calculate the variance of the CIF estimate at each time point $t_p$ as: \begin{equation*}\label{eq:EEAlgIF2}
\begin{split}
\hat{V}(\hat{F}_k^a(t_p)) = \frac{v(t_p; \tilde{D}_{S})}{N^2}
\end{split}
\end{equation*}
\end{algorithm*}

\subsubsection{Inference on the Federated CIF}

The influence function is also similar to the score function of a parametric estimator in the sense that the mean square value of the influence function at any particular time $t$ is a consistent estimator for the variance of the CIF estimate:
\begin{equation}\label{eq:inference}
\begin{split}
\sqrt{N}\left(\hat{F}_k^a(t) - F_k^a(t)\right) \xrightarrow{d} N(0, E[\psi_k^a(t,o_i)^2]).
\end{split}
\end{equation}
To conduct inference, we can express this in terms of the influence function values at each site:
\begin{equation}\label{eq:matrix}
\begin{split}
\widehat{\text{Var}}(\hat{F}_k^a(t)) &= 
\frac{\sum_{s=1}^{S}\sum_{D_s} \psi_k^a(\hat{F}_{k}^a(t); o_i)^2}{N^2}
\end{split}
\end{equation}
The numerator in \eqref{eq:matrix} can be computed as a cumulative sum across sites, and is safely shareable given the number of time points $Q<<N$. This allows us to compute confidence intervals at a prespecified set of time points. 

As previously noted, we treat the weights as known when estimating the empirical influence function.  This is consistent with existing non-parametric survival approaches that use the influence function, such as the targeted maximum likelihood approach described by \cite{Rytgaard2024}. Other widely used non-federated approaches also treat the weights as known, such as the IPW-KM approach described by \cite{Xie2005} and the IPW competing risks method developed by \cite{Conner2021}. Somewhat counter-intuitively, treating the weights as known typically leads to over-estimation of variance \citep{Austin2016, Chen2025}, rather than under-estimation, and has led practitioners to advocate use of the bootstrap in single-center studies to improve power. As the communication requirements to estimate bootstrap variance in a federated setting are unrealistic, we chose to treat the weights as known and accept slightly over-conservative inference as a limitation. This limitation may be addressed by deriving an influence function that includes additional augmentation terms to account for the impact of estimating the propensity score model.

\subsubsection{Federated RMTL}

A popular method for comparing CIF estimates between groups is to evaluate the difference in restricted mean time lost (RMTL), which we can estimate using our final curve:
\begin{equation}\label{eq:RMTL}
\begin{split} 
\hat{\mu}_1 - \hat{\mu}_0 = \int_0^{\tau} \hat{F}_k^1(t)dt - \int_0^{\tau}\hat{F}_k^0(t) dt
\end{split}
\end{equation}
RMTL measures the amount of survival time lost to cause $k$ up to a time $\tau$, and weighted RMTL is a preferred method to the Fine-Gray model when the proportional hazards assumption is violated \citep{Conner2021}. A negative value for the difference in RMTL between the treated counterfactual CIF and the untreated counterfactual CIF indicates that a treatment is protective against events due to cause $k$. The choice of time $\tau$ can pre-specified based on the clinical question of interest, but should be selected to ensure a sufficient number of patients remain in the risk set to ensure a stable estimates. If the choice of $\tau$ is unclear, our algorithm can calculate RMTL for multiple potential choices in parallel.  

The variance of the RMTL is typically estimated using the bootstrap, which is extremely difficult to execute in a federated context due to the intensive communication  and site-level expertise required. Instead we note that the RMTL has influence function $\omega_a^k(o_i)$ equal to the integral of the influence function of the CIF estimate:  
\begin{equation}\label{eq:RMTLIF}
\begin{split} 
\omega_a^k(o_i) = \int_0^{\tau} \psi_k^a(t,o_i) dt
\end{split}
\end{equation}
As each site already needs to calculate CIF influence function values for all local observations, they can easily estimate this quantity for each observation, allowing the variance of the RMTL difference to be estimated, again using a running sum shared across sites: 
\begin{equation}\label{eq:RMTLVAR}
\begin{split} 
\text{Var}(\hat{\mu}_1 - \hat{\mu}_0) = \frac{1}{N_{t}^2}\sum_{\{i:A_i=1\}} \hat{\omega}_1^k(o_i)^2 + \frac{1}{N_{c}^2}\sum_{\{i:A_i=0\}} \hat{\omega}_0^k(o_i)^2
\end{split}
\end{equation}
where $N_t$ is the total treated sample size and $N_c$ the total control sample size.

\section{Simulations}
\label{sec:sim}

We evaluated our method across several simulation scenarios considering two competing events, a main event of interest and a competing ``death" event. Each simulation scenario emulates an observational study where patients are followed for 25 months, with approximately 50\% of patients experiencing the main event of interest and 35\% of patients dying by month 15. $N=2000$ patients were distributed across 12 sites, with site-specific sample sizes ranging from 5 to 1000. In each simulation, we generated two normally distributed confounders associated with both the outcome and the binary exposure of interest. We designed three different simulation scenarios, denoted \textbf{A}, \textbf{B}, \textbf{C} in increasing order of complexity. In scenario \textbf{A}, failure times were exponentially distributed (constant event rate) with a 35\% probability of censoring at month 15. In scenarios \textbf{B} and \textbf{C}, failure times were Weibull distributed with shape parameter $\kappa = 0.5$ (higher event rate at the beginning of follow-up) with a censoring probability of 35\% in scenario \textbf{B} and 50\% in scenario \textbf{C}. In each of the scenarios, we considered a null setting, where the exposure did not impact the main event of interest, and two alternative settings where the exposure was positively associated with increased risk of the main event of interest (HR=$\{1.2,1.4\}$). Note that this induces a true negative association between the exposure and death, because the two event types compete with each other. For each simulation scenario, we conducted 500 experiment repeats. We compared our IP-weighted method to the meta-analysis method and the corresponding pooled results (gold standard) in terms of estimating difference in RMTL up to 25 months between the exposed and unexposed group. The meta-analysis estimates were obtain by fitting a local propensity score model, obtaining a CIF estimate, and calculating RMTL within each individual site, then pooling estimates based on inverse variance weighting.  For the meta-analysis method, we removed sites where the local sample size was too small to fit the propensity score model

\begin{figure*}[h]
\centering
{\includegraphics[width=\textwidth]{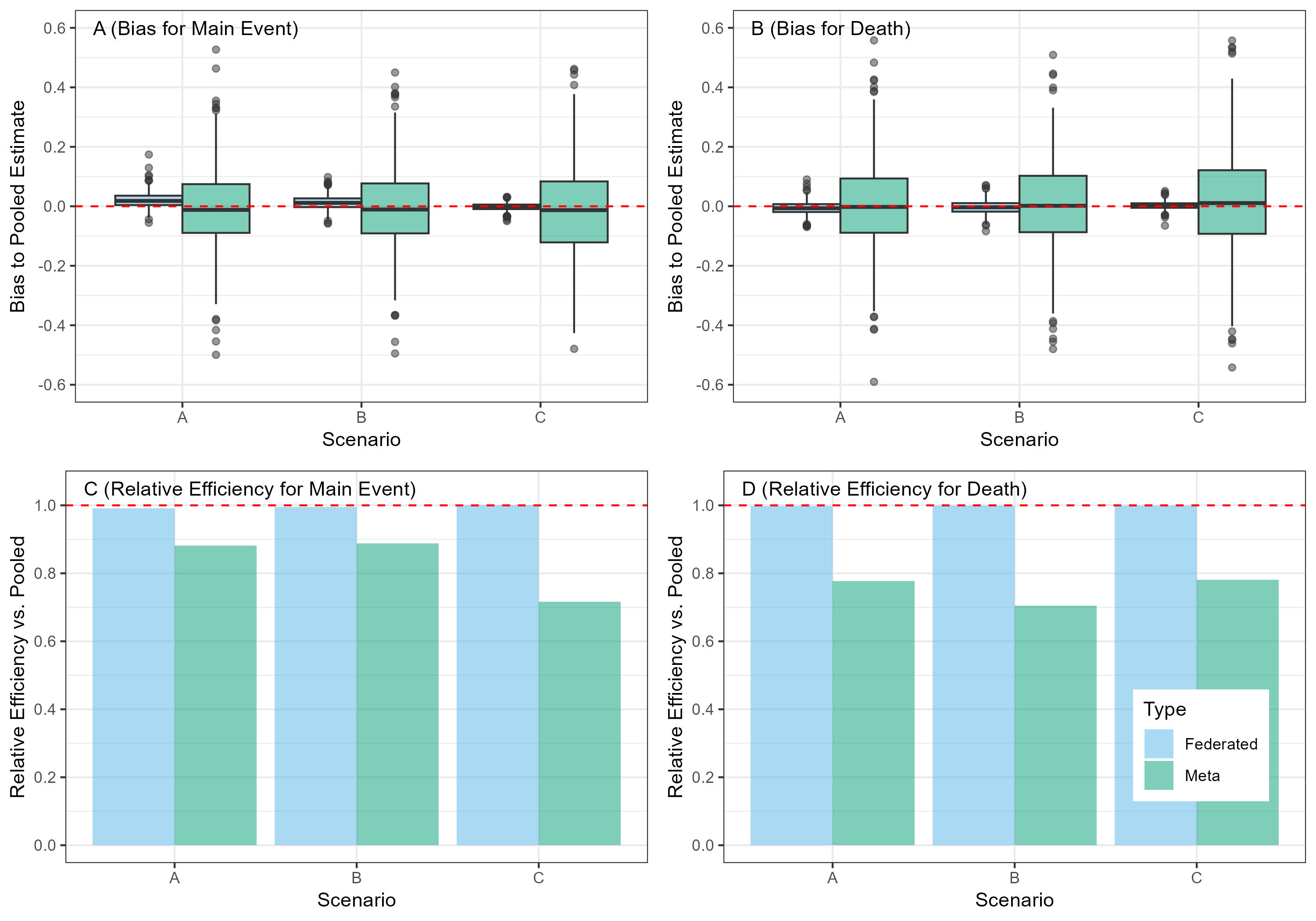}}
\caption{Comparison of meta-analysis and federated methods for distribution of bias to the pooled estimate (gold standard) for the main event of interest (panel A), and death (panel B), and relative efficiency compared to the pooled estimate for the main event (panel C) and death (panel D) across 500 simulation repeats for difference in RMTL between exposure and control groups, for simulation scenarios A (exponential failure times), B (Weibull failure times), C (Weibull failure times, high censoring). Abbrevs: RMTL; restricted mean time lost.\label{fig:second}}
\end{figure*}

\begin{figure*}[h]
\centering
{\includegraphics[width=\textwidth]{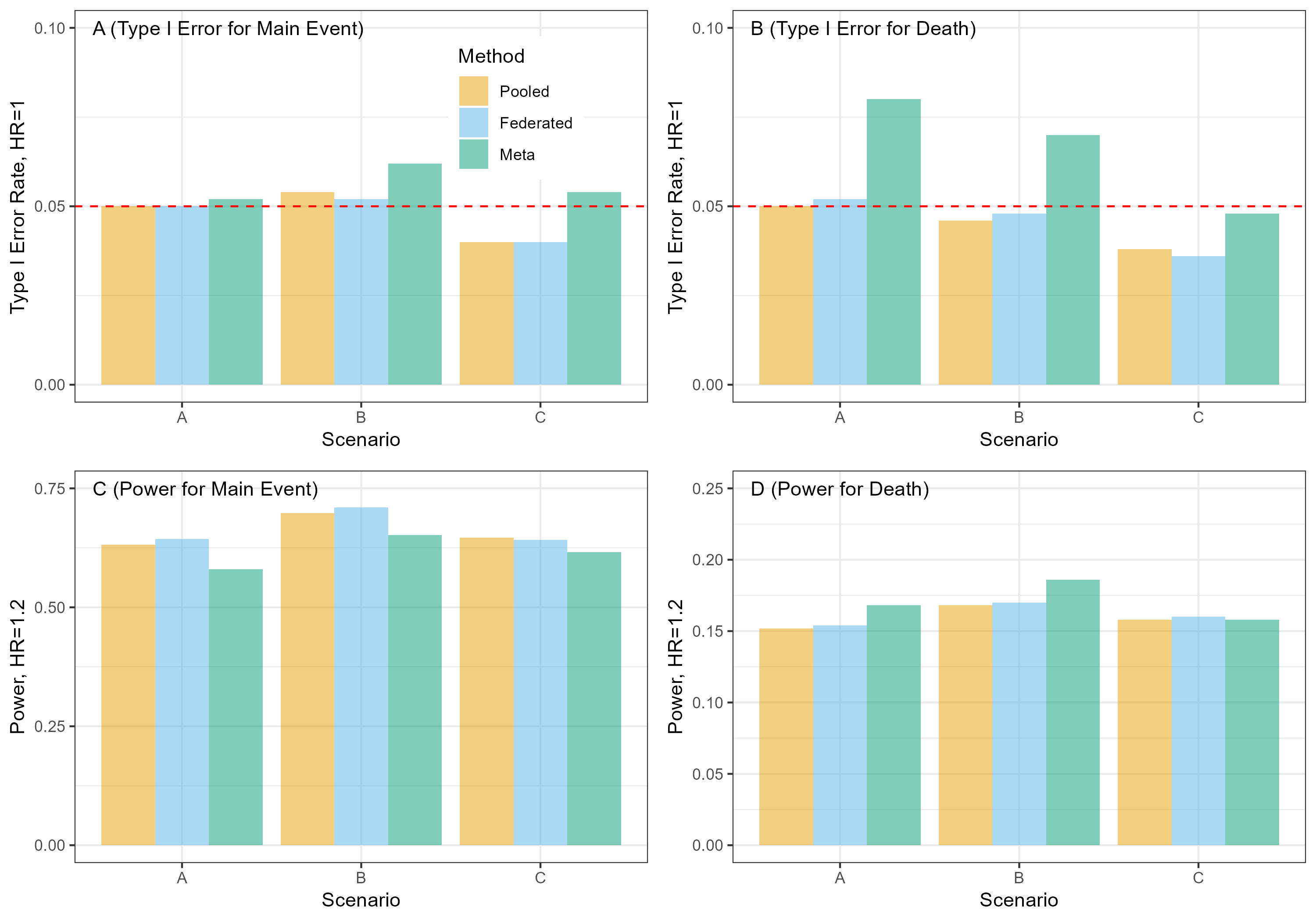}}
\caption{Type I error rate for the main event (panel A), and death (panel B), and power under HR=1.2 for the main event (panel C) and death (panel D) across 500 simulation repeats for difference in RMTL, for simulation scenarios A (exponential failure times), B (Weibull failure times), C (Weibull failure times, high censoring). Abbrevs: RMSE; root mean squared error. RMTL; restricted mean time lost.\label{fig:third}}
\end{figure*}

Figure \ref{fig:second} shows bias to the pooled estimator (gold standard) and relative efficiency when estimating the difference in RMTL between exposed and control groups, for both the main event of interest and competing event (``death"). Our method had negligible bias and equivalent efficiency to the pooled estimator, while meta-analysis was far less efficient. This difference was especially marked for the high censoring scenario, which corresponds to many real-world applications where we want to study rare event types and exposures. Figure \ref{fig:third} shows type I error rate and power for difference in RMTL for the pooled analysis, our federated method, and meta-analysis. Our method had similar statistical properties to the pooled analysis across all scenarios, with type I error controlled at 5\% and similar power to the pooled analysis. In contrast, meta-analysis suffered from either inferior power or inflated type I error across all three scenarios.

\section{Application}
\label{sec:app}

We applied our method to study the incidence of new endocrine immune-related adverse events (irAEs) in cancer patients receiving immune-checkpoint inhibitors (ICIs) at sites in the OneFlorida+ research network. Our goal was to investigate whether pre-existing auto-immune disease (AID) that were not endocrine-related increased the incidence of new irAEs following treatment with ICIs, while accounting for competing risk of death. We excluded patients who had a pre-existing endocrine-related AID as we would expect endocrine irAEs in this population regardless of treatment with ICIs. Based on time from first ICI given to either death or irAE (whichever occurred first), we used our previously published federated Kaplan-Meier method to construct cumulative incidence curves for the composite event of death or irAE for patients with and without pre-existing AID. We assessed difference between the curves based on RMST up to 18 months. We conducted an unweighted analysis and an IP-weighted analysis controlling for drug type (PDL, PD1, or CTLA4-inhibitor), age, gender, and Charlson Comorbidity Index (CCI). We then used the method developed in this paper to construct cumulative incidence curves for time to irAE while accounting for competing risk of death, again for patients with and without pre-existing AID. We assessed differences between the curves based on RMTL up to 18 months, which we selected as the maximum time where at least 25\% of the population remained at-risk in both exposure groups. We again conducted an unweighted analysis, and a weighted analysis controlling for the aforementioned variables. In all analyses, normalized weights were $<= 10$, indicating no instability concerns from extreme weight values.

There were $N=10,281$ patients receiving ICIs across $K=10$ sites, with site-specific sample sizes ranging from $n_1 = 3495$ to $n_{10}=1$. $929$ patients, or $9\%$, had a pre-existing non-endocrine AID. Our cohort was mostly male ($6168$, 60\%), older (65+: $5038$, 49\%), receiving PD1-inhibitors ($8019$, 78\%), and highly comorbid (CCI 5+: $5572$, 54\%). 

In the unweighted analysis, patients with pre-existing AID had lower irAE-free survival with an 18-month RMST of 10.7 months (95\% CI: [10.2,11.1]) compared to an 18-month RMST of 12.1 months (95\% CI: [11.9, 12.2]) for patients without pre-existing AID. Our competing risk analysis revealed that this was driven by a much higher incidence of irAEs, with patients in the pre-existing AID group losing a mean of 4.6 months (95\% CI: [4.2,5.1]) of irAE-free time prior to 18 months compared to 3.2 months (95\% CI: [3.0,3.3]) in the group without prior AID.

\begin{figure*}[h]
\centering
{\includegraphics[width=\textwidth]{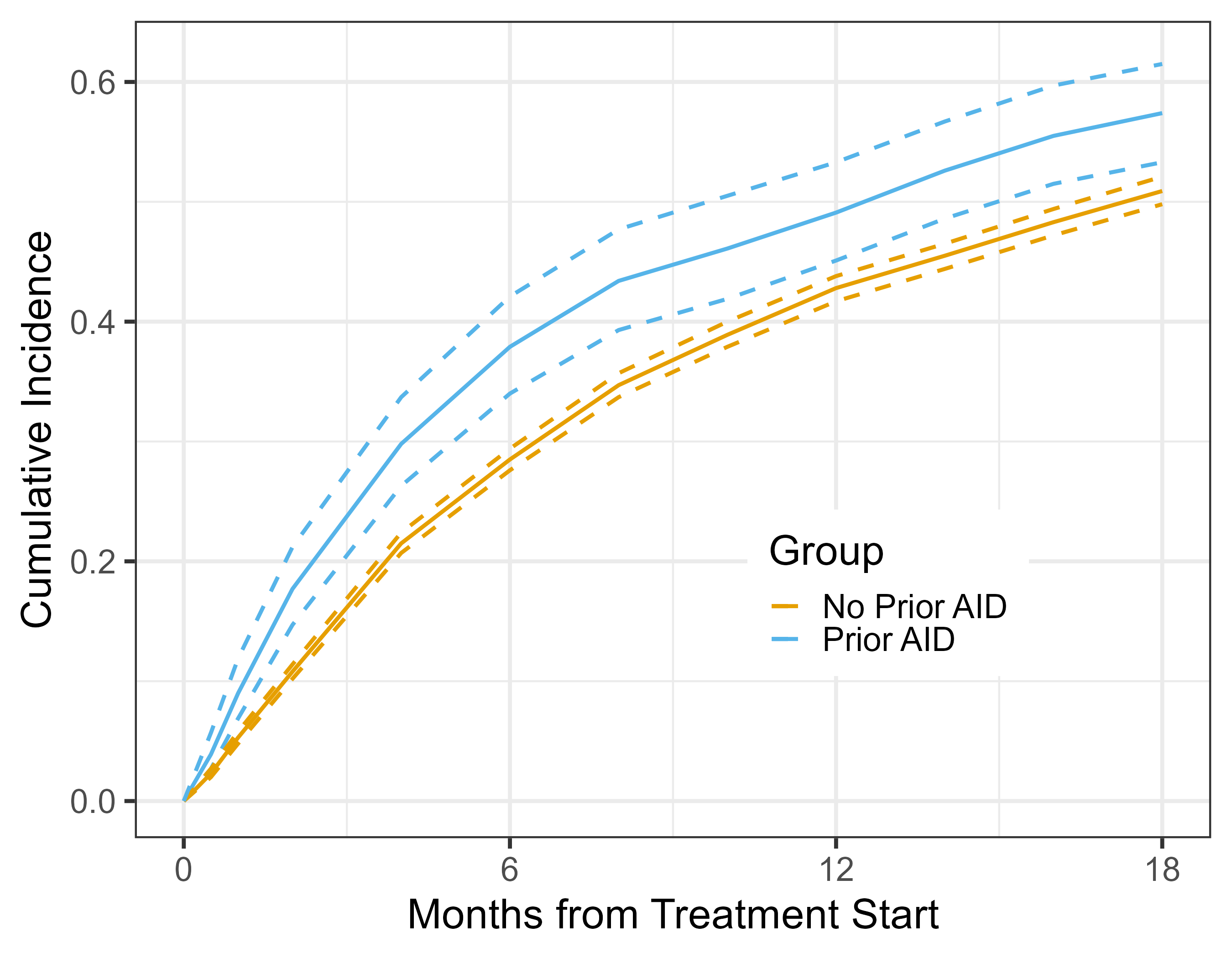}}
\caption{Weighted cumulative incidence of endocrine irAE or death for patients with and without pre-existing AID. Abbrevs: irAEs; immune-related adverse events, AID; auto-immune disease.\label{fig:EFS}}
\end{figure*}

\begin{figure*}[h!]
\centering
{\includegraphics[width=\textwidth]{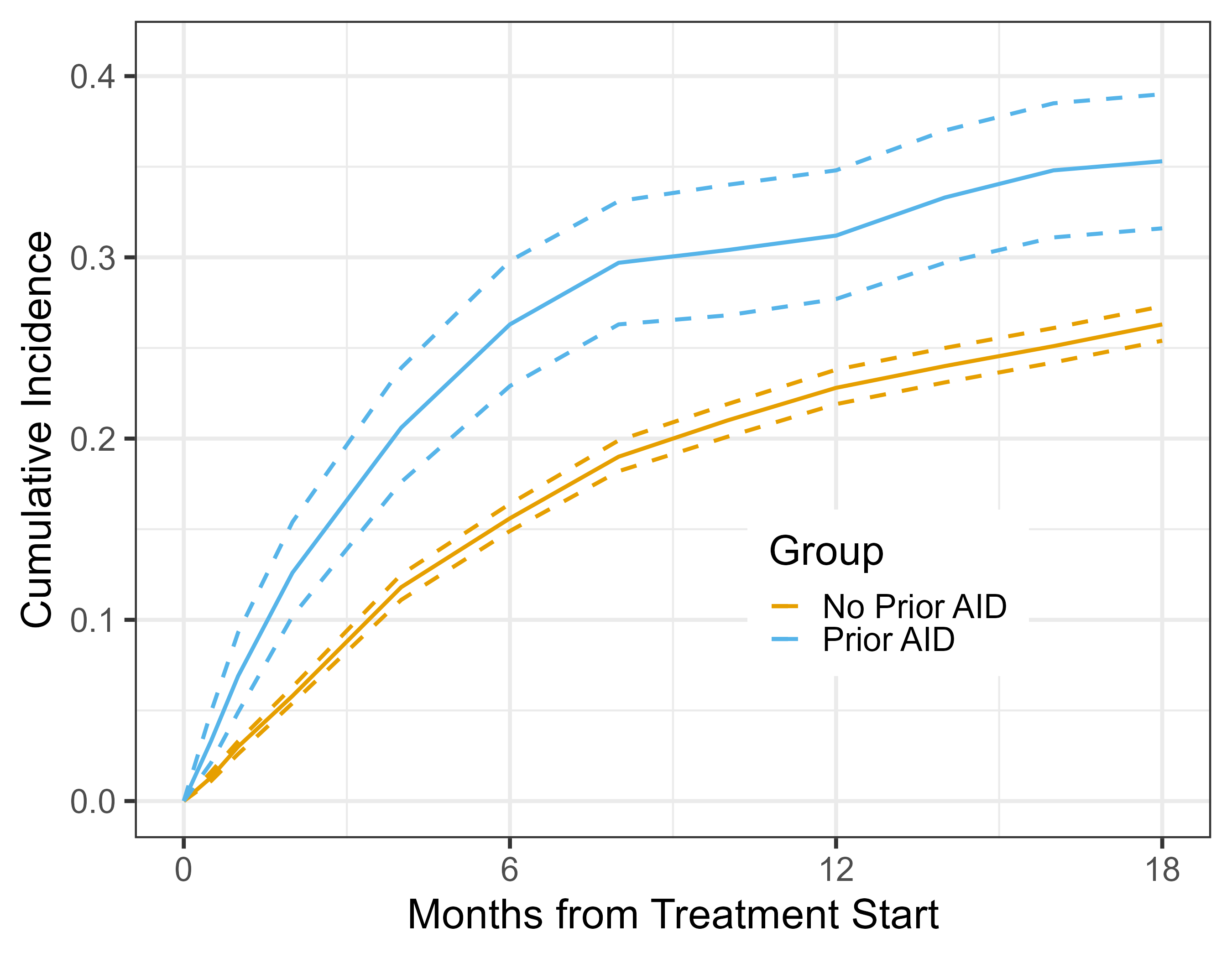}}
\caption{Weighted cumulative incidence of endocrine irAEs for patients with and without pre-existing AID. Abbrevs: irAEs; immune-related adverse events, AID; auto-immune disease.\label{fig:IRAE}}
\end{figure*}

The weighted analysis showed similar results. Patients with pre-existing AID had lower weighted irAE-free survival with an 18-month RMST of 10.8 months (95\% CI: [10.2,11.3]) compared to an 18-month RMST of 12.1 months (95\% CI: [11.9, 12.2]) for patients without pre-existing AID (see Figure \ref{fig:EFS}). Our competing risk analysis (see Figure \ref{fig:IRAE}) revealed that this was driven by a much higher weighted incidence of irAEs, with patients in the pre-existing AID group losing a mean of 4.8 months (95\% CI: [4.3,5.2]) of irAE-free time prior to 18 months compared to 3.2 months (95\% CI: [3.1,3.3]) in the group without prior AID.

\section{Conclusion}
\label{sec:conc}

We present a novel federated method for conducting competing risks analysis. Considering the impact of competing risks and choosing an appropriate analysis method is an increasingly important aspect of clinical trials and observational studies \citep{Andersen2012, Villanueva2022}. Our method provides crucial value relative to existing approaches because it avoids the proportional hazards assumption and allows for informative data visualization of covariate-adjusted treatment-stratified event incidence. Existing federated regression approaches require researchers to accept assumptions that are rarely true in practice \citep{Trinquart2016}, or go through a difficult-to-coordinate exploratory analysis and model fitting process across multiple sites to adequately account for time-varying effects. By providing researchers with a simple-to-implement federated method for constructing adjusted cumulative incidence curves, we both make the process of assessing proportional hazards assumptions much easier and provide a superior approach where these assumptions are violated. We also provide easily interpretable output quantities in the form of event probabilities and restricted mean time lost, which are easily communicated to clinicians and patients to provide a tangible picture of treatment effectiveness or exposure-related risk. 

We demonstrated our approach by conducting a multi-center study of irAEs following treatment with immune-checkpoint inhibitors in the OneFlorida+ research network. We found that patients with a prior non-endocrine auto-immune disease were more likely to experience endocrine-related irAEs following ICI treatment compared to control patients. This finding agrees with prior work \citep{Capelli2022, Haanen2020, Lusa2022}, and is one of the largest studies conducted of irAEs in this patient population. The estimates of cumulative event rate and irAE-free time lost provided in this study can help clinicians and patients make informed decisions, and allow for better monitoring of adverse events in this high-risk population. 

As with other analysis methods, such as the Fine-Gray model, cumulative incidence of events in a competing risk setting needs to be carefully interpreted. As individuals who experience a competing event remain in the risk set, a treatment can sometimes appear to have a protective effect simply by increasing the incidence of competing events \citep{Young2020}. Adequately characterizing the impact of treatments requires distinguishing between the \textit{direct} effect of treatment on the event of interest, and \textit{indirect} effect through increasing or decreasing incidence of competing events. A limitation of our method is that the difference in RMTL can only represent the combination of these two effects, or the \textit{total} effect of exposure on the event of interest. We can study the direct effect of treatment using the more traditional Kaplan-Meier method along with inverse probability of censoring weights (IPCW) to control for the impact of censoring by competing events. Extending our previously developed federated methods for KM analysis \citep{Risk2025} and restricted mean survival time to allow for IPCW is a potential future direction which would address this limitation and allow for a more fulsome treatment of competing risks data. 

\bigskip
\begin{center}
{\large\bf SUPPLEMENTARY MATERIAL}
\end{center}

\begin{description}

\item[Supplement:] Technical appendix including theoretical details (A), github link to R packages along with detailed user instructions (B). (pdf)

\end{description}

\bigskip
\begin{center}
{\large\bf ACKNOWLEDGMENTS AND FUNDING}
\end{center}

Research reported in this publication was supported by the National Institute Of Allergy And Infectious Diseases of the National Institutes of Health under Award Number R01AI158543. The content is solely the responsibility of the authors and does not necessarily represent the official views of the National Institutes of Health. 

\bigskip
\begin{center}
{\large\bf DISCLOSURE STATEMENT}
\end{center}

The authors report there are no competing interests to declare.

\bigskip
\begin{center}
{\large\bf DATA AVAILABILITY}
\end{center}

Data are not available without approval from the University of Florida. Data inquiries can be addressed to the corresponding author.

\bibliographystyle{agsm}

\bibliography{JBI_REFS}
\end{document}

\title{Technical Appendix for Federated Competing Risks Analysis}
\maketitle
\section{Appendix A: Theoretical Results}

\subsection{The Influence Function (Kaplan-Meier)} 

To describe the influence function of the KM estimator, we start by defining the generating distribution of the censored data pairs $(X, \Delta)$ in terms of the cumulative subdistribution function $G$:
\begin{equation*}
\begin{split}
G(t, \delta)=\mathbb{P}(X < t, \Delta = \delta)
\end{split}
\end{equation*}
The true survival function $S$ admits representation as a functional $\Psi(G)$:
\begin{equation*}
\begin{split}
S(t) = \Psi(G)(t) = \exp\left(-\int_0^{t}\frac{dG(s,0)}{G(s,0)+G(s,1)}\right)
\end{split}
\end{equation*}
Peterson showed in 1977 that the KM estimate $\hat{S}$ is the result of applying $\Psi$ to the empirical cumulative subdistribution function of the data $\hat{G}$: 
\begin{equation*}
\begin{split}
\hat{S}(t) = \Psi(\hat{G})(t) = \exp\left(-\int_0^{t}\frac{d\hat{G}(s,0)}{\hat{G}(s,0)+\hat{G}(s,1)}\right)
\end{split}
\end{equation*}
To compute the influence function, we apply a small perturbation to G, based on a distribution placing all of it's mass at a single observation $(X_i, \Delta_i)$:
\begin{equation*}
\begin{split}
G_{\epsilon}(t, \delta) = (1-\epsilon)\mathbb{P}(X < t, \Delta = \delta) + \epsilon I(\Delta_i = \delta)I(X_i < t)
\end{split}
\end{equation*}
Then the influence function of $S$ is given by:
\begin{equation*}
\begin{split}
\psi(X_i, \Delta_i)(t) = \left.\frac{\partial \Psi(G_{\epsilon})(t)}{\partial \epsilon}\right|_{\epsilon=0}
\end{split}
\end{equation*}
As shown in Reid (1981), following this equation through will give the expression:
\begin{equation*}\label{eq:IFKM_Reid}
\psi(S; X_i,\Delta_i)(t)= -S(t)\left[\frac{\Delta_i I(X_i\leq t)}{Y(X_i)} -\int_0^{\text{min}(X_i, t)} \frac{\lambda(u)}{Y(u)}du \right]         \end{equation*}
Given a known weight $w_i$, the equivalent expression for the counterfactual survival function $S^a(t)$ under treatment level $a$ is given by:
\begin{equation*}\label{eq:INFKM}
\begin{split}
\psi(S^a; X_i,\Delta_i, w_i)(t)= -w_iS^a(t)\left[\frac{\Delta_i I(X_i\leq t)}{Y^a(X_i)} -\int_0^{\text{min}(X_i, t)} \frac{\lambda^a(u)}{Y^a(u)}du \right] 
\end{split}
\end{equation*}

\subsection{Influence Function (Competing Risks)}

The cumulative incidence for competing risk $j$ can be defined in the following way:
\begin{equation*}
\begin{split}  
F_k(t) = \int_0^t \lambda_k(s)S(s)ds  = \int_0^t \lambda_k(s) \exp\left(-\int_0^s \sum_{k=1}^K \lambda_k(u)du\right)ds
\end{split}
\end{equation*}
where here $S$ represents overall survival from all $K$ competing risks. Writing in terms of subdistribution hazard functions as in section 1.1 we have: 
\begin{equation*}
\begin{split} 
F_k(t) = - \int_0^t S(s) \frac{dS^{u,k}(s)}{Y(s)} 
\end{split}
\end{equation*}
where:
\begin{equation*}
\begin{split} 
S^{u,k}(t) &= \mathcal{P}(X_i > t, \Delta_i=k) \\
Y(t) &= \mathcal{P}(X_i > t)
\end{split}
\end{equation*}
for event indicator $\Delta_i \in \{0,...,K\}$ and follow-up time $X_i$. Then the influence function $\psi(F_k; O)(t)$ will be given by:
\begin{equation*}
\begin{split} 
\psi(F_k; O)(t) &= \left.\frac{d}{d\epsilon} F^{(\epsilon)}_k(t)\right|_{\epsilon=0}\\
F_k^{(\epsilon)}(t) &= -\int_0^t S^{(\epsilon)}(s)\frac{dS^{u,k,(\epsilon)}(s)}{Y^{(\epsilon)}(s)} \\
S^{u,k,(\epsilon)}(t) &= (1-\epsilon)\mathcal{P}(X_i > t, \Delta_i=k)+\epsilon I(X_i > t, \Delta_i=k) \\
Y^{(\epsilon)}(t) &=(1-\epsilon) \mathcal{P}(X_i > t) + \epsilon I(X_i > t) \\
-dS^{u,k,(\epsilon)}(t) &= f^{u,k,(\epsilon)}(t) = (1-\epsilon) f^{u,k}(t) + \epsilon \mathbb{I}_{X_i}(t) I(\Delta_i = k) 
\end{split}
\end{equation*}
Now by product rule:
\begin{equation*}
\begin{split} 
 \left.\frac{d}{d\epsilon} F^{(\epsilon)}_k(t)\right|_{\epsilon=0} &= -\int_0^t \left. \frac{dS^{(\epsilon)}(s)}{d\epsilon}\frac{dS^{u,k,(\epsilon)}(s)}{Y^{(\epsilon)}(s)}\right|_{\epsilon=0} + \left.S^{(\epsilon)}(s) \frac{d}{d\epsilon}\frac{dS^{u,k,(\epsilon)}(s)}{Y^{(\epsilon)}(s)}\right|_{\epsilon=0} \\ &= -\int_0^t \psi(S; o_i)(s) \left. \frac{dS^{u,k,(\epsilon)}(s)}{Y^{(\epsilon)}(s)}\right|_{\epsilon=0} + \left.S^{(\epsilon)}(s) \frac{d}{d\epsilon}\frac{dS^{u,k,(\epsilon)}(s)}{Y^{(\epsilon)}(s)}\right|_{\epsilon=0} 
\end{split}
\end{equation*}
where above $\psi(S, s)(o_i)$ represents the influence function of the ordinary KM estimator. Then following through this expression eventually gives us: 
\begin{equation*}
\begin{split} 
\psi(F_k; o_i)(t) = \frac{I(\Delta_i=k) I(X_i \leq t) S(X_i)}{Y(X_i)} + \int_0^t\lambda_k(s)\psi(S; o_i)(s)ds - \int_0^{\min(t, X_i)} \frac{S(s)\lambda_k(s)}{Y(s)}ds
\end{split}
\end{equation*}

Given a known weight $w_i$, the equivalent expression for the counterfactual cumulative incidence function $F^a_k(t)$ under treatment level $a$ is given by:
\begin{equation*}\label{eq:INFCR}
\begin{split}
\psi(F^a_k;o_i)(t) = \frac{w_iI(\Delta_i=k) I(X_i \leq t) S^a(X_i)}{Y^a(X_i)} + \int_0^t\lambda_k^a(s)\phi(S^a;o_i)(s)ds- w_i\int_0^{t \wedge X_i} \frac{S^a(s)\lambda_k^a(s)}{Y^a(s)}ds
\end{split}
\end{equation*}

\cleardoublepage

\subsection{Update Algorithm}

For illustrative purposes, we derive the algorithm under randomized conditions, i.e. true weight $w_i = 1$ for all $i$. The proof given varying propensity scores is similar. In order to derive the update algorithm for $\hat{F}^a_k(t)$, we will require the following result: \par \bigskip

\textbf{Lemma.} The derivative $\dot{\psi}(F^a_k(t);o_i)$ of the estimated influence function $\psi(\hat{F}^a_k(t);o_i)$ at time $t$ satisfies the following:
\begin{equation*}
\begin{split}
\frac{1}{N} \sum_{i=1}^N \dot{\psi}(\hat{F}^a_k(t);o_i) = \frac{1}{N} \sum_{i=1}^N  \frac{\partial \psi(\hat{F}^a_k(t);o_i)}{\partial \hat{F}^a_k(t)} \overset{p}{\to} -1
\end{split}
\end{equation*}

\textit{Proof of Lemma.} Recall our expression from the main text, substituting $\hat{w}_i = 1$:
\begin{equation*}\label{eq:INF}
\begin{split}
\psi(\hat{F}^a_k(t);o_i) &= \frac{I(\Delta_i=k) I(X_i \leq t) \hat{S}^a(X_i)}{\hat{Y}^a(X_i)} + \sum_{t_j \leq t}\frac{\hat{F}^a_k(t_j)- \hat{F}^a_k(t_{j-1})}{\hat{S}^a(t_{j-1})}\phi(\hat{S}^a(t_j);o_i)\\&- \sum_{t_j \leq (t \wedge X_i)} \frac{\hat{F}^a_k(t_j)- \hat{F}^a_k(t_{j-1})}{\hat{Y}^a(t_j)}
\end{split}
\end{equation*}
Taking the derivative, the first term disappears and we retain only the last term in the sum for each of the second and third terms: 
\begin{equation*}\label{eq:INF}
\begin{split}
\dot{\psi}(\hat{F}^a_k(t);o_i) =\frac{\partial \psi(\hat{F}^a_k(t);o_i)}{\partial \hat{F}^a_k(t)} = \frac{\phi(\hat{S}^a(t);o_i)}{\hat{S}^a(t)}- \frac{I(X_i > t)}{\hat{Y}^a(t)}
\end{split}
\end{equation*}

Now note that $\frac{1}{N} \sum_{i=1}^N \frac{\phi(\hat{S}^a(t);o_i)}{\hat{S}^a(t)} = \frac{1}{N\hat{S}^a(t)} \sum_{i=1}^N \phi(\hat{S}^a(t);o_i) \overset{p}{\to} 0$ by the weak law of large numbers as the influence function has mean zero and finite variance. We also know that $\hat{Y}^a(t) = \frac{1}{N} \sum_{i=1}^N I(X_i > t)$ and hence:
\begin{equation*}\label{eq:conv}
\begin{split}
\frac{1}{N} \sum_{i=1}^N \dot{\psi}(\hat{F}^a_k(t);o_i) \overset{p}{\to} 0 - \frac{1}{N\hat{Y}^a(t)} \sum_{i=1}^N I(X_i > t) = - \frac{\hat{Y}^a(t)}{\hat{Y}^a(t)} = -1,
\end{split}
\end{equation*}
as required. \begin{flushright} $\blacksquare$ \end{flushright}

To obtain the updating algorithm, consider the case where we have incorporated data from the first $s-1$ sites $\tilde{D}_{s-1}$ into our estimate $\hat{F}^a_k(t;\tilde{D}_{s-1})$ and want to add data from a new site $D_{s}$. We know that the combined data estimate $\hat{F}^a_k(t;\tilde{D}_{s})$ will satisfy the following score equation:
\begin{equation*}
\begin{split}
\sum_{\tilde{D}_{s}} \psi(\hat{F}^a_k(t); o_i) &= 0 \\
\Rightarrow \sum_{\tilde{D}_{s-1}} \psi(\hat{F}^a_k(t); o_i) + \sum_{D_{s}} \psi(\hat{F}^a_k(t); o_i) &= 0 \\
\end{split}
\end{equation*}

Since we cannot calculate $\sum_{\tilde{D}_{s-1}} \psi(\hat{F}^a_k(t); o_i)$ but have access to the current estimate $\hat{F}^a_k(t;\tilde{D}_{s-1})$, we consider the Von Mises expansion (see Gill 1989) of $\hat{F}^a_k(t;\tilde{D}_{s-1})$:
\begin{equation*}
\begin{split}
\hat{F}^a_k(t;\tilde{D}_{s-1}) = F^a_k(t) + \frac{1}{\tilde{N}_{s-1}}\sum_{\tilde{D}_{s-1}} \psi(F^a_k(t); o_i)  + o_p(n^{-1/2})
\end{split}
\end{equation*}

For any estimator $\tilde{F}^a_k$ near the true value $F^a_k(t)$ we can take a Taylor expansion of the second term:
\begin{equation*}
\begin{split}
\hat{F}^a_k(t;\tilde{D}_{s-1}) = F^a_k(t) + \frac{1}{\tilde{N}_{s-1}}\sum_{\tilde{D}_{s-1}} \psi(\tilde{F}^a_k(t); o_i)+ \left[\frac{1}{\tilde{N}_{s-1}}\sum_{\tilde{D}_{s-1}} \dot{\psi}(\tilde{F}^a_k(t); o_i)\right](\tilde{F}^a_k(t) - F^a_k(t))   + o_p(n^{-1/2})
\end{split}
\end{equation*}
Using the lemma from the beginning of this section, we then have the approximation:
\begin{equation*}
\begin{split}
\hat{F}^a_k(t;\tilde{D}_{s-1}) &\approx F^a_k(t) + \frac{1}{\tilde{N}_{s-1}}\sum_{\tilde{D}_{s-1}} \psi(\tilde{F}^a_k(t); o_i)+ \left[-1\right](\tilde{F}^a_k(t) - F^a_k(t)) \\ &= \tilde{F}^a_k(t) + \frac{1}{\tilde{N}_{s-1}}\sum_{\tilde{D}_{s-1}} \psi(\tilde{F}^a_k(t); o_i) 
\end{split}
\end{equation*}
For any consistent estimator $\tilde{F}^a_k(t)$, this gives us the approximation:
\begin{equation*}
\begin{split}
\tilde{N}_{s-1} [\hat{F}^a_k(t;\tilde{D}_{s-1}) - \tilde{F}^a_k(t)] \approx \frac{1}{\tilde{N}_{s-1}}\sum_{\tilde{D}_{s-1}} \psi(\tilde{F}^a_k(t); o_i)
\end{split}
\end{equation*}
This means we can approximate the first term in our score without access to previous data:
\begin{equation*}
\begin{split}
\sum_{\tilde{D}_{s-1}} \psi(\hat{F}^a_k(t); o_i) + \sum_{D_{s}} \psi(\hat{F}^a_k(t); o_i) \approx \tilde{N}_{s-1} [\hat{F}^a_k(t;\tilde{D}_{s-1}) - \hat{F}^a_k(t)] + \sum_{D_{s}} \psi(\hat{F}^a_k(t); o_i) 
\end{split}
\end{equation*}
Then again using our lemma, we can solve this approximate score equation using Newton's method:
\begin{equation*}\label{eq:EEAlg}
\begin{split}
\hat{F}_k^a(t)^{(r+1)} = \hat{F}^a_k(t)^{(r)} + \frac{\tilde{N}_{s-1} [\hat{F}^a_k(t; \tilde{D}_{s-1})- \hat{F}^a_k(t)^{(r)}] + \sum_{D_{s}} \psi({\hat{F}_{k}^a(t)}^{(r)};o_i)}{\tilde{N}_{s-1} +N_s}
\end{split}
\end{equation*}
where the denominator above is asymptotically equivalent to the derivative of the score equation from the property $\frac{1}{N} \sum_{i=1}^N \dot{\psi}(\hat{F}^a_k(t);o_i) \overset{p}{\to} -1$. 

\section{Appendix B: Technical Note on Algorithm}

\subsection{Installation}

All R functions required to implement the distributed method, along with detailed documentation, can be found at https://github.com/MR236/FederatedCR.\par

\subsection{Knot Selection}

We recommend selecting knot locations at equally spaced quantiles of the survival distribution as observed at the first site (as in the default settings), and using 4-10 knots with spline degree of at least 2. If the first site is large, the number of knots desired can be used at the first site. If the first site is smaller, we would recommend adding an additional knot for every 100-200 observations at subsequent sites until the target is reached. \par

\section{References}

Hines O., Dukes O., Diaz-Ordaz K. and Vansteelandt S. (2022). Demystifying Statistical Learning Based on Efficient Influence Functions. \textit{The American Statistician}, \textbf{76(3),} 292-304. \bigskip

Gill, R. D., Wellner, J. A., and Præstgaard, J. (1989). Non- and Semi-Parametric Maximum Likelihood Estimators and the Von Mises Method (Part 1) [with Discussion and Reply]. \textit{Scandinavian Journal of Statistics}, \textbf{16(2),} 97–128.\bigskip

James, L. F. (1997). A study of a class of weighted bootstraps for censored data. \textit{Annals of Statistics}, \textbf{25,} 1595–1621.\bigskip

Peterson, A.V. (1977). Expressing the Kaplan-Meier estimator as a function of empirical sub-survival functions. \textit{JASA}, \textbf{72,} 854-858.\bigskip

Reid, N. (1981). Influence Functions for Censored Data. \textit{Ann. Statist.}, \textbf{9(1)} 78-92.